\documentclass{ws-rv9x6}
\usepackage[square]{ws-rv-van}             
\makeindex
\begin{document}

\chapter[Finite Temperature Matrix Product State Algorithms and Applications]{Finite Temperature Matrix Product State Algorithms and Applications \label{wall_mpsthermal}}

\author[M. L. Wall and L. D. Carr]{Michael L. Wall and Lincoln D. Carr}

\address{Department of Physics, Colorado School of Mines,\\
Golden, Colorado 80401, USA \\
mwall@mines.edu}

\begin{abstract}
We review the basic theory of matrix product states (MPS) as a numerical variational ansatz for time evolution, and present two methods to simulate finite temperature systems with MPS: the ancilla method and the minimally entangled typical thermal state method.  A sample calculation with the Bose-Hubbard model is provided.\\
\end{abstract}

\body

\section{Introduction }

The dimension of the Hilbert space for a general many-body system increases exponentially with the system size, severely restricting the sizes which are amenable to straightforward numerical study.  Several techniques have been developed to deal with this fact, such as the stochastic sampling of the Hilbert space in quantum Monte Carlo and the judicious use of symmetries and sparse matrix structures in exact diagonalizations.  The most successful approximate method for 1d systems is the density matrix renormalization group (DMRG) method first pioneered by White \cite{white_92}.  Soon after, the theory of matrix product states\cite{klumper_schadschneider_91 ,fannes_nachtergaele_92} (MPS) was used to shed light on the amazing success of DMRG\cite{oestlund_rommer_95, oestlund_rommer_97}.  Ideas from quantum information theory, most notably the idea of bipartite entanglement, have led to the development of MPS algorithms which generalize DMRG to time evolution\cite{vidal_04, daley_kollath_04}, periodic boundary conditions \cite{verstraete_porras_04}, and finite temperature\cite{verstraete_garciaripoll_04, zwolak_vidal_04}.  A thorough discussion of the time-evolving block decimation algorithm, an MPS algorithm for zero temperature time evolution, is given in chapter \ref{daley_mpstebd }.  In this chapter we review algorithms based on MPS for finite temperature simulations and discuss their relevance to studying finite temperature superfluid systems.

\section{Methodology} 
\subsection{Matrix Product States}
\label{wall::sec:MPS}
A \emph{matrix product state}\footnote{An MPS is a vector in Hilbert space.  The qualifier matrix product refers to the fact that the expansion coefficients in the Fock basis are expressed as products of matrices.} (MPS) is defined as
\begin{equation}
|\Psi_{\mathrm{mps}}\rangle=\sum_{i_1,i_2,\dots i_L=1}^{d}\mathrm{Tr}\left(\rm{\bf{A}}^{\left[1\right]i_1}\cdots \rm{\bf{A}}^{\left[L\right] i_L}\right)|i_1,\cdots , i_L\rangle
\end{equation}
where the $\rm{\bf{A}}^{\left[k\right] i_k}$ are matrices\footnote{These matrices can be taken to have the same symmetry as the state they represent, e.g., if the state has real coefficients in some basis then the MPS matrices can be taken to be real.  See \cite{perezgarcia_verstraete_07} and references therein for more details.} the dimension of which is a fixed number $\chi$ known as the bond dimension, $d$ is the dimension of the Hilbert space spanned by the $\left\{|i_k\rangle \right\}$, and $L$ is the number of lattice sites.  Let us refer to the set of all MPSs with bond dimension $\chi$ as $\mathcal{M}_{\chi}$.  An MPS in $\mathcal{M}_{\chi}$ contains $Ld\chi^2$ parameters, and so it is clear that any state on a finite lattice can be written as an MPS provided we take the bond dimension to be $\chi_{\mathrm{max}}=d^{\lfloor L/2\rfloor}$.  However, the great utility of MPSs is that an MPS with bond dimension $\chi\ll \chi_{\mathrm{max}}$ often provides an excellent approximation to the true state\cite{verstraete_cirac_06} and allows for exponentially more efficient manipulation and calculation of observables than an exact representation.
\begin{figure}[htbp]
\centerline{\includegraphics[width=4.25cm]{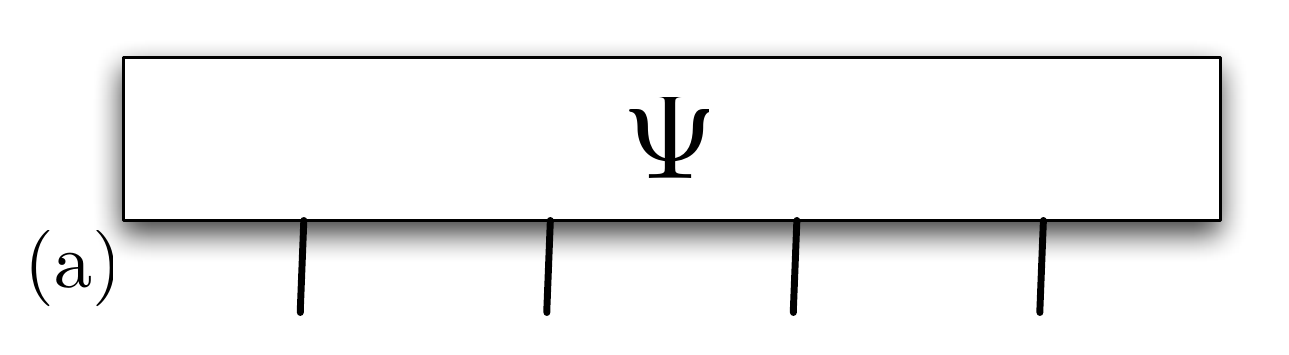}\includegraphics[width=4.25cm]{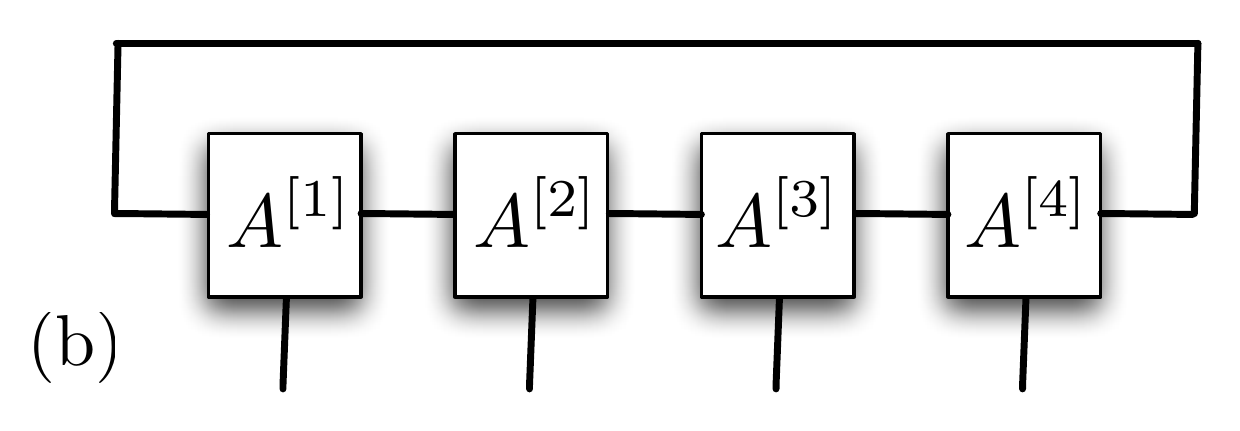}}
\centerline{\includegraphics[width=4.25cm]{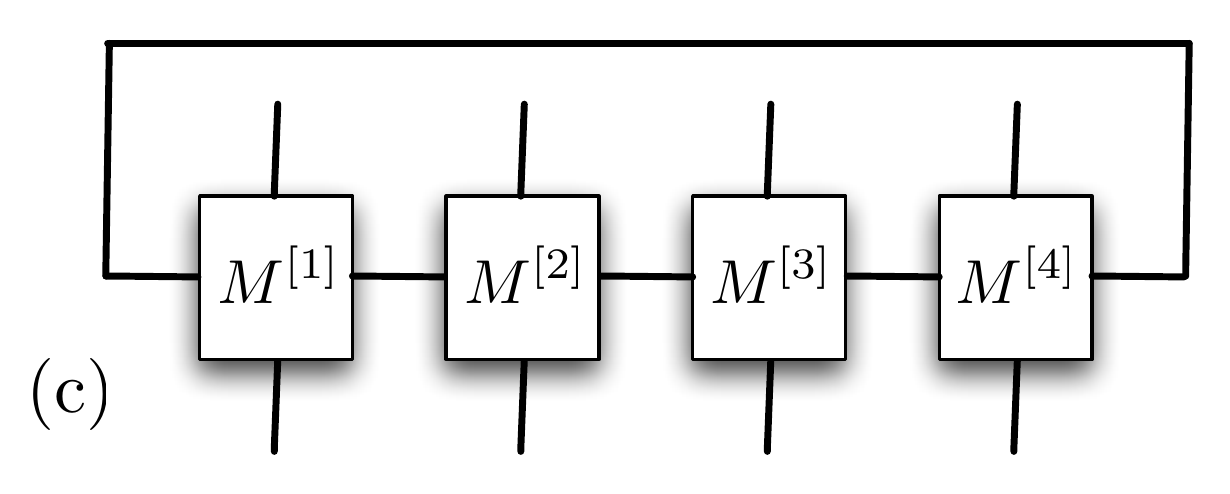}}
\caption{ (a) Tensor network representation of full 4 site wavefunction.  (b) Tensor network representation of an MPS on 4 sites.  (c) Tensor network representation of an MPO on 4 sites.}
\label{wall:fig:tensornetworkdiagram}
\end{figure}

To visualize MPSs and operations with them,  it is useful to introduce the notion of a \emph{tensor network diagram} as in Fig.~\ref{wall:fig:tensornetworkdiagram}.  In such a diagram a box represents a tensor, free lines are uncontracted indices and closed lines are contracted indices.  Fig.~\ref{wall:fig:tensornetworkdiagram}(a) shows the state of a many-body system expressed in the basis of the full Hilbert space as an  $L$-index  tensor, and Fig.~\ref{wall:fig:tensornetworkdiagram}(b) shows the same state written as an MPS.  The advantage of the MPS representation becomes clear when we compute scalar products such as $\langle \psi|\hat{O}|\phi\rangle$.

Before we discuss how scalar products are efficiently computed, it is advantageous to introduce a \emph{matrix product operator} (MPO) as 
\begin{equation*}
\hat{O}=\sum_{i_1,\dots, i_L=1}^{d}\sum_{i_1',\dots ,i_L'=1}^{d}\mathrm{Tr}\left(\rm{\bf{M}}^{\left[1\right] i_1i_1'}\cdots \rm{\bf{M}}^{\left[L\right] i_Li_L'}\right)|i_1,\cdots , i_L\rangle\langle i_1',\cdots ,i_L'|\, ,
\end{equation*}
where each of the $\rm{\bf{M}}^{\left[k\right] i_k i_k'}$ is a matrix the dimensions of which are bounded by a fixed number $\chi_O$ known as the bond dimension.  The tensor network representation of an MPO is similar to that of an MPS, but there are two uncontracted indices per tensor corresponding to the bra and ket indices; see Fig.~\ref{wall:fig:tensornetworkdiagram}(c).  Equivalently, one can think of each element of the matrix $M^{\left[k\right]}$ as being operator valued, where the operator acts on the space spanned by $\left\{|i_k\rangle\right\}$.

\begin{figure}
\includegraphics[width=11cm]{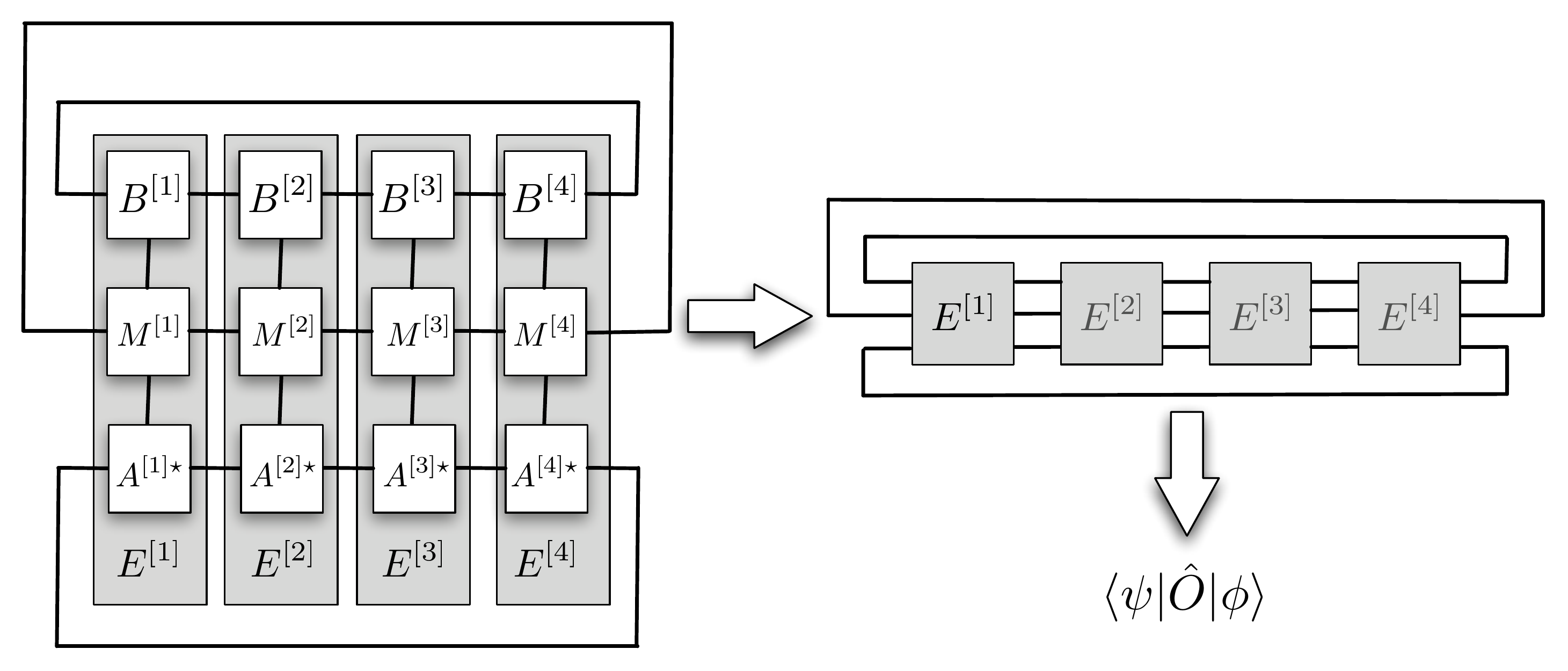}
\caption{Tensor network representation of the scalar product procedure Eqs.~\eqref{wall::Eq:scProd}-\eqref{wall::Eq:transmat}.  The transfer matrices $\rm{\bf{E}}^{\left[k\right]}_{\rm{\bf{M}}}\left(\rm{\bf{A}},\rm{\bf{B}}\right)$ have been abbreviated as $E^{\left[k\right]}$ for succinctness.}
\label{wall:fig:scProd}
\end{figure}

Let us now see how to evaluate the scalar product of an operator $\hat{O}$ represented as an MPO between two states $|\psi\rangle$ and $|\phi\rangle$ represented as MPSs.  Let us denote the MPO matrices of $\hat{O}$ as $\rm{\bf{M}}$ and the MPS matrices of $|\psi\rangle$ and $|\phi\rangle$ as $\rm{\bf{A}}$ and $\rm{\bf{B}}$, respectively.  Then, we have
\begin{eqnarray}
\nonumber \langle \psi |\hat{O}|\phi\rangle&=&\sum_{i_1,\dots, i_L=1}^{d}\sum_{i_1',\dots, i_L'=1}^{d}\mathrm{Tr}\left({\rm{\bf{A}}^{\left[1\right]i_1}}^{\star}\cdots {\rm{\bf{A}}^{\left[L\right]i_L}}^{\star}\right)\mathrm{Tr}\left(\rm{\bf{M}}^{\left[1\right]i_1i_1'}\cdots \rm{\bf{M}}^{\left[L\right] i_Li_L'}\right)\\
 &&\times\mathrm{Tr}\left(\rm{\bf{B}}^{\left[1\right]i_1}\cdots \rm{\bf{B}}^{\left[L\right]i_L}\right)\\
\label{wall::Eq:scProd} \nonumber&=&\mathrm{Tr}\Big(\left[\sum_{i_1,i_1'=1}^{d}{\rm{\bf{A}}^{\left[1\right] i_1}}^{\star}\otimes \rm{\bf{M}}^{\left[1\right] i_1i_1'}\otimes \rm{\bf{B}}^{\left[1\right]i_1'}\right]\times \cdots\\
&&\times  \left[\sum_{i_L,i_L'=1}^{d}{\rm{\bf{A}}^{\left[L\right] i_L}}^{\star}\otimes \rm{\bf{M}}^{\left[L\right] i_Li_L'}\otimes \rm{\bf{B}}^{\left[L\right]i_L'}\right]\Big)\\
\label{wall::Eq:transmat}&\equiv&\mathrm{Tr}\left(\rm{\bf{E}}^{\left[1\right]}_{\rm{\bf{M}}}\left(\rm{\bf{A}},\rm{\bf{B}}\right)\cdots \rm{\bf{E}}^{\left[L\right]}_{\rm{\bf{M}}}\left(\rm{\bf{A}},\rm{\bf{B}}\right)\right)\, ,
\end{eqnarray}
where the last line defines the generalized transfer matrix $\rm{\bf{E}}^{\left[k\right]}_{\rm{\bf{M}}}\left(\rm{\bf{A}}, \rm{\bf{B}}\right)\equiv \sum_{i_k,i_k'=1}^{d}{\rm{\bf{A}}^{\left[k\right] i_k}}^{\star}\otimes \rm{\bf{M}}^{\left[k\right] i_ki_k'}\otimes \rm{\bf{B}}^{\left[k\right]i_k'}$, which is a $\chi^2\chi_O\times \chi^2\chi_O$ matrix.  Naively we would expect that the multiplication of two transfer matrices would require $\mathcal{O}\left(\chi^6\chi_O^3\right)$ operations, but the special structure of the transfer matrices allows us to perform such a multiplication in $\mathcal{O}\left(\chi^5\chi_O^2d^2\right)$\footnote{The fact that the boundary matrices of MPSs with open boundary conditions have bond dimension 1 allows us to perform this contraction in $\mathcal{O}\left(\chi^3\chi_O^2d^2\right)$, and recent developments for periodic boundary conditions have reduced the scaling to $\mathcal{O}\left(\chi^3\chi_O^2d^2\right)$ for large systems with only a few relevant correlation lengths\cite{pippan_white_10,pirvu_verstraete_10}.}  as
\begin{eqnarray}
&&\left[\rm{\bf{E}}^{\left[k\right]}_{\rm{\bf{M}}}\left(\rm{\bf{A}},\rm{\bf{B}}\right)\rm{\bf{E}}^{\left[k+1\right]}_{\rm{\bf{M}}}\left(\rm{\bf{A}},\rm{\bf{B}}\right)\right]_{\left[\alpha \gamma \beta\right],\left[\alpha'\gamma'\beta'\right]}\\
\nonumber &&=\sum_{i'=1}^{d}\sum_{\beta=1}^{\chi}\left(\sum_{i=1}^{d}\sum_{\gamma''=1}^{\chi_O}\left(\left[\rm{\bf{G}}_{\rm{\bf{M}}}^{\left[k\right]}\left(\rm{\bf{A}},\rm{\bf{B}}\right)\right]_{\left[\alpha \gamma \beta\right],\left[\alpha' \gamma'' \beta'\right]}\right)\rm{\bf{M}}_{\gamma''\gamma'}^{\left[k+1\right]ii'}\right)\rm{\bf{B}}^{\left[k+1\right]i'}_{\beta''\beta'} \\
\nonumber&&\left[\rm{\bf{G}}_{\rm{\bf{M}}}^{\left[k\right]}\left(\rm{\bf{A}},\rm{\bf{B}}\right)\right]_{\left[\alpha \gamma \beta\right],\left[\alpha' \gamma'' \beta'\right]}\equiv\sum_{\alpha''=1}^{\chi}\left[\rm{\bf{E}}^{\left[k\right]}_{\rm{\bf{M}}}\left(\rm{\bf{A}},\rm{\bf{B}}\right)\right]_{\left[\alpha \gamma \beta\right],\left[\alpha'' p\gamma''\beta''\right]}{\rm{\bf{A}}^{\left[k+1\right]i}_{\alpha'' \alpha'}}^{\star}\, .
\end{eqnarray}
Here the square brackets around indices denote a composite index in the Kronecker representation and parentheses give the order in which the contraction should be performed to ensure the best scaling.  In particular, it is essential not to sum over the $\alpha''$ and $\beta''$ indices simultaneously.\footnote{Here and throughout we use greek indices to denote bond indices and roman indices to denote physical indices.}  The tensor network representation of the scalar product procedure is given in Fig.~\ref{wall:fig:scProd}.

Many operators of interest, such as translation invariant 1d Hamiltonians, can be easily represented as MPOs with small bond dimension $\chi_O\sim$4-10\cite{mcculloch_07,pirvu_murg_10}, and the MPO representations of more complex operators can be constructed using simple MPO arithmetic~\cite{mcculloch_07, stoudenmire_white_10}.  That the MPO form of an operator is optimal for MPS algorithms can be straightforwardly deduced using the tensor network formalism, as the scalar product of an MPO between two MPSs is the most general tensor network that can be efficiently contracted; see Fig.~\ref{wall:fig:scProd}.

\begin{figure}
\centerline{\includegraphics[width=7cm]{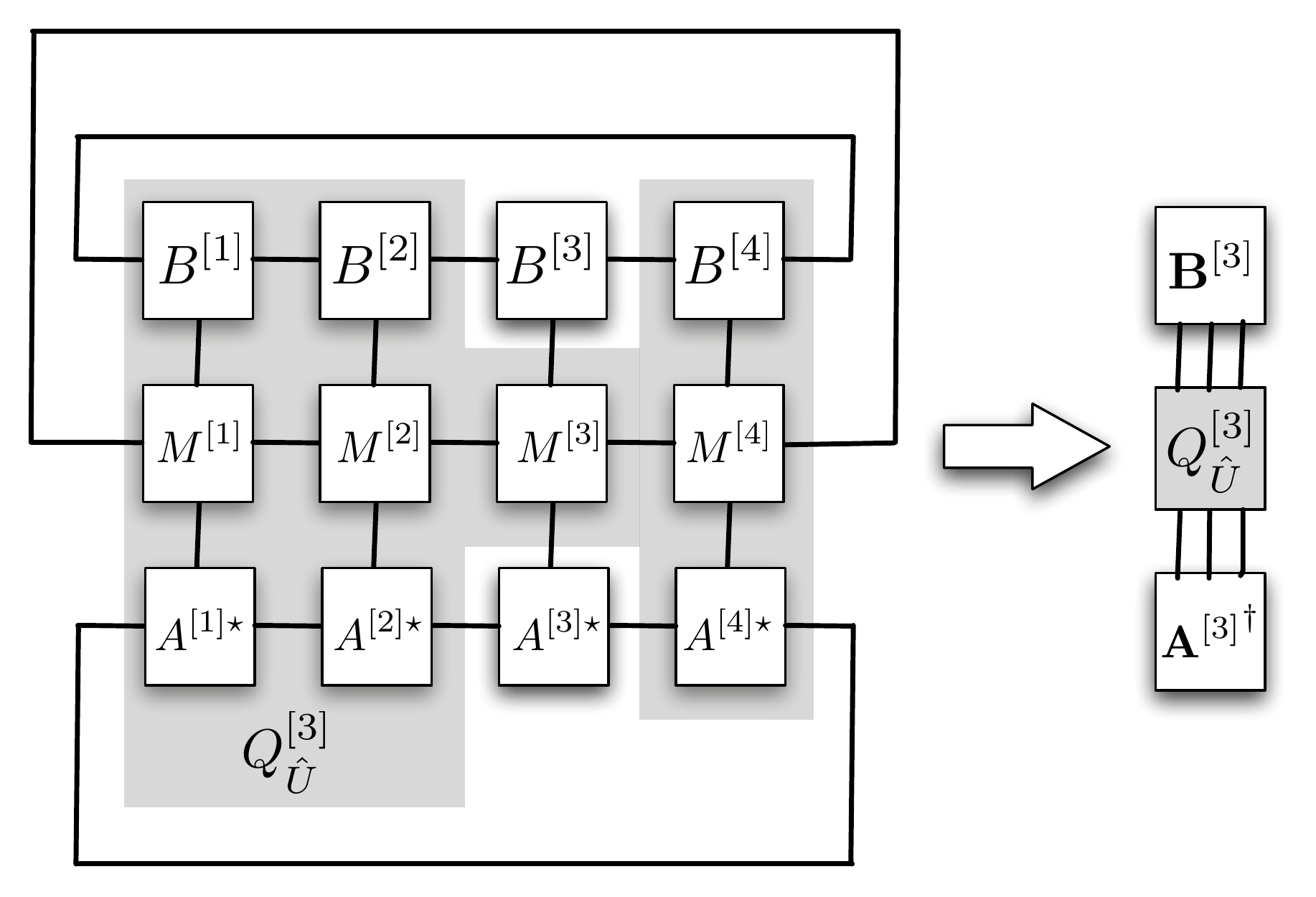}}
\caption{Tensor network representation of the quadratic form representing $\langle \psi|\hat{U}|\phi\rangle$ in Eq.~\eqref{wall::Eq:quad}.}
\label{wall:fig:quad}
\end{figure}

We now turn to the simulation of time evolution using MPSs.  The main difficulty of using MPSs is that $\mathcal{M}_{\chi}$ is not a vector space.\footnote{This can be seen from the fact that the addition of two MPSs is given by the direct sum of their matrices: $|\psi_{C}\rangle=|\psi_A\rangle+|\psi_B\rangle\Rightarrow \rm{\bf{C}}^{\left[k\right]}=\rm{\bf{A}}^{\left[k\right]}\oplus \rm{\bf{B}}^{\left[k\right]}$.  If the matrices $\rm{\bf{A}}^{\left[k\right]}$ and $\rm{\bf{B}}^{\left[k\right]}$ have orthogonal bases then $\dim\left(\rm{\bf{C}}^{\left[k\right]}\right)=\dim\left(\rm{\bf{A}}^{\left[k\right]}\right)+\dim\left(\rm{\bf{B}}^{\left[k\right]}\right)$.}  Thus, when operators such as the propagator are applied to an MPS we must find the optimal\footnote{By optimal we mean that the overlap is maximal in the 2-norm.  Although MPSs do not form a vector space, they are embedded in a larger Hilbert space and so this norm is well-defined.} projection into $\mathcal{M}_{\chi}$ to keep the algorithm efficient.  We denote this projection as $\mathcal{P}_{\chi}$.  The optimal MPS $|\psi\rangle\in\mathcal{M}_{\chi}$ representing the MPS $\hat{U}|\phi\rangle$ is
\begin{eqnarray}
\mathcal{P}_{\chi}\left[\hat{U}|\phi\rangle\right]&=&\min_{|\psi\rangle\in \mathcal{M}_{\chi}}\left||\psi\rangle-\hat{U}|\phi\rangle\right|^2\\
\label{wall::Eq:quad}&=&\min_{|\psi\rangle\in\mathcal{M}_{\chi}}\left[\langle \psi|\psi\rangle+\langle \phi|\hat{U}^{\dagger}\hat{U}|\phi\rangle-2\mathcal{R}\left(\langle \psi|\hat{U}|\phi\rangle\right)\right]\, ,
\end{eqnarray}
where $\mathcal{R}\left(\bullet\right)$ denotes the real part.  Each of the scalar products in Eq.~\eqref{wall::Eq:quad} may be written as a quadratic form in each of the matrices $\rm{\bf{A}}^{\left[k\right]i_k}$, as is demonstrated in the tensor network diagram Fig.~\ref{wall:fig:quad}. 
\noindent \\\\

Again denoting the matrices in the MPS representation of $|\psi\rangle$ by $\rm{\bf{A}}$ and those of $|\phi\rangle$ by $\rm{\bf{B}}$, the quadratic form of the $k^{\mathrm{th}}$ site may be written as
\begin{equation}
\label{wall::Eq:quad2} \mathcal{Q}^{\left[k\right]}={\mathbf{A}^{\left[k\right]}}^{\dagger}\rm{\bf{Q}}^{\left[k\right]}_{\hat{1}}\mathbf{A}^{\left[k\right]}+{\mathbf{B}^{\left[k\right]}}^{\dagger}\rm{\bf{Q}}^{\left[k\right]}_{\hat{U}^{\dagger}\hat{U}}\mathbf{B}^{\left[k\right]}-2\mathcal{R}\left({\mathbf{A}^{\left[k\right]}}^{\dagger}\rm{\bf{Q}}^{\left[k\right]}_{\hat{U}}\mathbf{B}^{\left[k\right]}\right)\, ,
\end{equation}
where $\mathbf{A}^{\left[k\right]}$ represents the $d\chi^2$ elements of the $\left\{\rm{\bf{A}}^{\left[k\right]i_k}\right\}$, arranged as a vector, and the matrices $\rm{\bf{Q}}_{\hat{O}}$ are defined as
\begin{equation}
\left[\rm{\bf{Q}}^{\left[k\right]}_{\hat{O}}\right]_{\left[\alpha i_k\alpha'\right]\left[\beta i_k'\beta'\right]}=\sum_{\gamma,\gamma'=1}^{\chi_O}\rm{\bf{M}}_{\gamma\gamma'}^{\left[k\right]i_ki_k'}\left[\prod_{j\ne k} \rm{\bf{E}}^{\left[j\right]}_{\rm{\bf{M}}}\left(\rm{\bf{C}},\rm{\bf{D}}\right)\right]_{\left[\alpha \gamma \beta \right],\left[\alpha' \gamma' \beta'\right]}\, ,
\end{equation}
where $\rm{\bf{C}}$ and $\rm{\bf{D}}$ are either $\rm{\bf{A}}$ or $\rm{\bf{B}}$ depending on the quadratic form.  The $\rm{\bf{M}}^{i_ki_k'}$ in this final expression are the matrices in the MPO representation of $\hat{O}$.  The stationary points of the quadratic form Eq.~\eqref{wall::Eq:quad2} are given by the solution of the linear system\footnote{It is important to note that while $\rm{\bf{Q}}_{\hat{1}}$ is the quadratic form representing the scalar product $\langle \psi|\psi\rangle$ it can not in general be made equal to the identity.  The numerical conditioning of this matrix and of the linear system Eq.~\eqref{wall::Eq:stationary} can be improved by suitable choice of ``gauge conditions" on the matrices $A$; see\cite{verstraete_porras_04}.}\cite{verstraete_porras_04}
\begin{equation}
\label{wall::Eq:stationary}\rm{\bf{Q}}_{\hat{1}}\mathbf{A}^{\left[k\right]}=\rm{\bf{Q}}_{\hat{U}}\mathbf{B}^{\left[k\right]}\, .
\end{equation}
The algorithmic procedure for time evolution is to sweep back and forth through the lattice, solving Eq.~\eqref{wall::Eq:stationary} at each site until convergence is reached.  In practice, it is essential for efficiency not to explicitly form the matrices $\rm{\bf{Q}}_{\bullet}$, but rather to use iterative methods which require only multiplication by the $\rm{\bf{Q}}_{\bullet}$ to solve Eq.~\eqref{wall::Eq:stationary}.  Details on the form of the propagator $\hat{U}$ can be found in \cite{garciaripoll_06, stoudenmire_white_10}.

\subsection{The Ancilla Method}
At finite temperature, the state of a quantum system is given by the thermal density matrix $\hat{\rho}=e^{-\beta\hat{H}}/Z$.  The ancilla method\cite{verstraete_garciaripoll_04, feiguin_white_05} relies on the notion of purification\cite{nielsen_chuang_book_00} to represent the thermal density matrix as a pure state in an enlarged Hilbert space.  Each physical site is augmented with an ancilla which has the same Hilbert space dimension as the physical site.  The MPS representation of such a system is
\begin{equation}
\label{wall::Eq:MPA}|\psi\rangle=\sum_{i_1,\dots , i_L=1}^{d}\sum_{a_1,\dots , a_L=1}^{d} \mathrm{Tr}\left(\rm{\bf{A}}^{\left[1\right] i_1a_1}\cdots \rm{\bf{A}}^{\left[L\right] i_La_L}\right)|i_1a_1\cdots i_La_L\rangle\, .
\end{equation}
One can think of the combined system as a two-legged ladder, with the physical sites on the lower leg and the ancillae on the upper leg.  The purpose of the ancillae is to act as a perfect heat bath which, when traced out, provides the proper thermal density matrix for the physical system.  The choice of ancilla for infinite temperature ($\beta=0$) is simply the normalized purification of the identity
\begin{equation}
\textstyle|\psi\left(0\right)\rangle=\frac{1}{\sqrt{d^L}}\prod_{k=1}^{L}\sum_{i_k,a_k=1}^{d}\delta_{i_ka_k}|i_ka_k\rangle\, ,
\end{equation}
which represents a product of maximally entangled site-ancilla pairs.  This state has an MPS representation with bond dimension 1 generated by taking all matrices to be $\rm{\bf{A}}^{\left[k\right] i_k a_k}_{\alpha\beta}=\delta_{\alpha,1}\delta_{\beta,1}\delta_{i_ka_k}/\sqrt{d}$.  The extension to finite inverse temperature $\beta$ is provided by evolving only the physical sites\footnote{That is, the Hamiltonian only couples physical sites to physical sites, and not ancillae to ancillae or physical sites to ancillae.} in imaginary time up to $\beta/2$,
\begin{equation}
|\psi\left(\beta\right)\rangle=e^{-\beta \hat{H}/2}|\psi\left(0\right)\rangle\,.
\end{equation}
This time evolution can be efficiently performed using the methods of Sec.~\ref{wall::sec:MPS}.  Observables are calculated using transfer matrices as above with the additional requirement that the ancilla degrees of freedom are traced over.

The ancilla method is conceptually very simple, and becomes numerically exact for large enough bond dimension.  However, because the MPS Eq.~\eqref{wall::Eq:MPA} has to encode the information of both the system and the bath, it requires a bond dimension $\sim\chi_{\mathrm{g.s.}}^2$ at low temperatures, where $\chi_{\mathrm{g.s.}}$ is the bond dimension required to accurately represent the ground state.  Typical values of $\chi_{\mathrm{g.s.}}$ range from 50-5000, making the ancilla method impractical for many systems at very low temperatures.

We conclude this section by remarking that the ancilla method represents a highly idealized heat bath chosen to reproduce the exact thermal density matrix.  Many of the current examples of strongly correlated many-body systems, e.~g.~cold atoms, are very mesoscopic and are in contact with thermal reservoirs which are also mesoscopic.  A modification of the ancilla method where the perfect entanglement at infinite temperature is replaced with ancilla-ancilla and ancilla-system couplings in the Hamiltonian can be devised.  Alternatively, one can directly simulate master equations by considering matrix product density operators with optimal projections based on the Hilbert-Schmidt distance\cite{verstraete_porras_04} or matrix product decompositions of superkets with local projections\cite{zwolak_vidal_04}.

\subsection{Minimally Entangled Typical Thermal States}
A new method for finite temperature MPS simulations has recently been proposed by White \cite{white_09}.  The idea stems from the question ``What is a typical wave function of a quantum system at finite temperature?"  That is, if we are to measure a quantum system at finite temperature, what ``typical" pure states would we find, and with what probabilities?  It is clear from the basic tenets of statistical mechanics that any set of typical states $\left\{|\phi\left(i\right)\rangle\right\}$ must satisfy
\begin{equation}
\label{wall::Eq:typicality}\sum_iP\left(i\right)|\phi\left(i\right)\rangle\langle \phi\left(i\right)|=e^{-\beta\hat{H}}\, ,
\end{equation}
where $P\left(i\right)$ is the probability of measuring the system to be in state $|\phi\left(i\right)\rangle$, and so the expectation of an operator $\hat{A}$ at finite temperature may be written as
\begin{equation}
\label{wall::Eq:avg}\langle \hat{A}\rangle=\sum_i\frac{P\left(i\right)}{Z}\langle \phi\left(i\right)|\hat{A}|\phi\left(i\right)\rangle\, ,
\end{equation}
with $Z$ the partition function.  From Eq.~\eqref{wall::Eq:avg}, we see that we can calculate observables using an unweighted average of $\langle \phi\left(i\right)|\hat{A}|\phi\left(i\right)\rangle$ if we choose the $|\phi\left(i\right)\rangle$ at random according to their probabilities of being measured, $P\left(i\right)/Z$.  It is easy to generate states satisfying the typicality condition Eq.~\eqref{wall::Eq:typicality} simply by defining any orthonormal basis $\left\{|i\rangle\right\}$ and defining
\begin{equation}
|\phi\left(i\right)\rangle=\left[P\left(i\right)\right]^{-1/2}\exp\left(-\beta\hat{H}/2\right)|i\rangle\, ,\;P\left(i\right)=\langle i|\exp\left(-\beta \hat{H}\right)|i\rangle\, .
\end{equation}
We now use the freedom in the choice of the orthonormal basis $\left\{|i\rangle\right\}$ to generate typical states with the least amount of spatial entanglement, as these are the states which can be most efficiently represented as MPSs\cite{vidal_03, vidal_04}.  This amounts to taking the $\left\{|i\rangle\right\}$ to be classical product states (CPSs), $|i\rangle=\prod_{k=1}^{L}|i_k\rangle$, where $i_k$ labels the state of site $k$.  The set of $|\phi\left(i\right)\rangle$ obtained from this choice of $\left\{|i\rangle\right\}$ are called \emph{minimally entangled typical thermal states} (METTS).

The most efficient algorithmic procedure for generating thermal averages using METTS is as follows:
\begin{enumerate}
\item Choose a CPS $|i\rangle$ at random.
\item Evolve in imaginary time using the methods of Sec.\ref{wall::sec:MPS} to generate the METTS $|\phi\left(i\right)\rangle=\left[P\left(i\right)\right]^{-1/2}\exp\left(-\beta\hat{H}/2\right)|i\rangle$.
\item Compute observables of interest using this METTS and add to the running averages.
\item Randomly select a new CPS $|i'\rangle$ according to the probability $\left|\langle i'|\phi\left(i\right)\rangle\right|^2$.
\item Repeat from step 2 until converged.
\end{enumerate}
We see that the main loop of this algorithm closely resembles a Monte Carlo iteration with measurement taking the place of the usual configuration updates.  However, it does not rely on a rejection method to perform sampling, and so each METTS that is generated can be used to generate statistics.  In practice very few ($\sim100$) METTS suffice to obtain the total energy to a relative accuracy of $10^{-5}$.  For algorithmic details on how to perform the CPS selection to minimize correlations between successive METTS we refer the reader to \cite{stoudenmire_white_10}.

This METTS algorithm has many advantages over the ancilla method of the previous section.  As we do not have to encode the bath degrees of freedom in our MPS, the bond dimension  required to accurately represent each METTS ranges from 1 at infinite temperature to $\chi_{\mathrm{g.s.}}$ at very low temperatures.  This makes the METTS method more efficient than the ancilla method by a factor of $10^{3}-10^{10}$ for typical systems at very low temperatures.  Additionally, if the Hamiltonian of interest has a global symmetry then we can use the fact that the MPS matrices must transform irreducibly to speed up the calculation\cite{mcculloch_07} or find the thermal ensemble corresponding to a fixed quantum number (canonical ensemble).  This latter point is relevant to cold atom systems where the total number of atoms is held fixed.\footnote{The ancilla method can also be used to simulate systems in the canonical ensemble, but the process is complicated by the fact that we need the purification of the constrained infinite temperature density matrix.  This purification can be generated using a ground state DMRG-type calculation with a suitably chosen Hamiltonian\cite{feiguin_fiete_10}.  The Hamiltonian will contain artificial ancilla-ancilla and ancilla-physical site couplings which are typically highly nonlocal.}
\section{Validity issues }
It has been shown that MPSs can faithfully represent ground states of 1d gapped Hamiltonians with at most nearest neighbor interactions with a bond dimension which grows only polynomially in the system size\cite{verstraete_cirac_06}.  In higher dimensions this polynomial scaling gives way to an exponential scaling\cite{Liang_Pang_94}, but calculations on 2D systems of width 8-12 are still feasible\cite{white_scalapino_09}.  Generalizations of MPSs to higher dimensions exist, but are so far limited by poor polynomial scaling of tensor contractions\cite{verstraete_cirac_04, schuch_wolf_08, evenbly_vidal_09}.  Perhaps the most important quality of MPS methods as compared to other efficient many body methods, such as quantum Monte Carlo, is that MPS methods work equally well for fermionic or frustrated systems.  All of the methods presented here will work equally well for any 1d or quasi-1d physical system.

\section{Application: Specific Heat of the Hard-core Extended Bose-Hubbard Model }

As an example of how the above methods may be applied to study the behavior of a finite temperature superfluid system, we study the properties of the hard-core extended Bose-Hubbard model
\begin{equation}
\textstyle\hat{H}=-J\sum_{\langle i,j\rangle}\left(\hat{b}_i^{\dagger}\hat{b}_j+\mathrm{H.c.}\right)+V\sum_{\langle i,j\rangle}\hat{n}_i\hat{n}_j
\end{equation}
at half filling.  This model is known to have a superfluid phase in the XY universality class for $V<2J$.  In the below figure we show a typical thermodynamic quantity, the specific heat $C_V=\beta^2(\langle \hat{H}^2\rangle-\langle\hat{H}\rangle^2)/L$, as a function of temperature and the nearest-neighbor repulsion.  Note that computation of $\langle \hat{H}^2\rangle$ is easily performed when the MPO representation of $\hat{H}$ is known.

\begin{figure}
\centerline{\includegraphics[width=8.0cm]{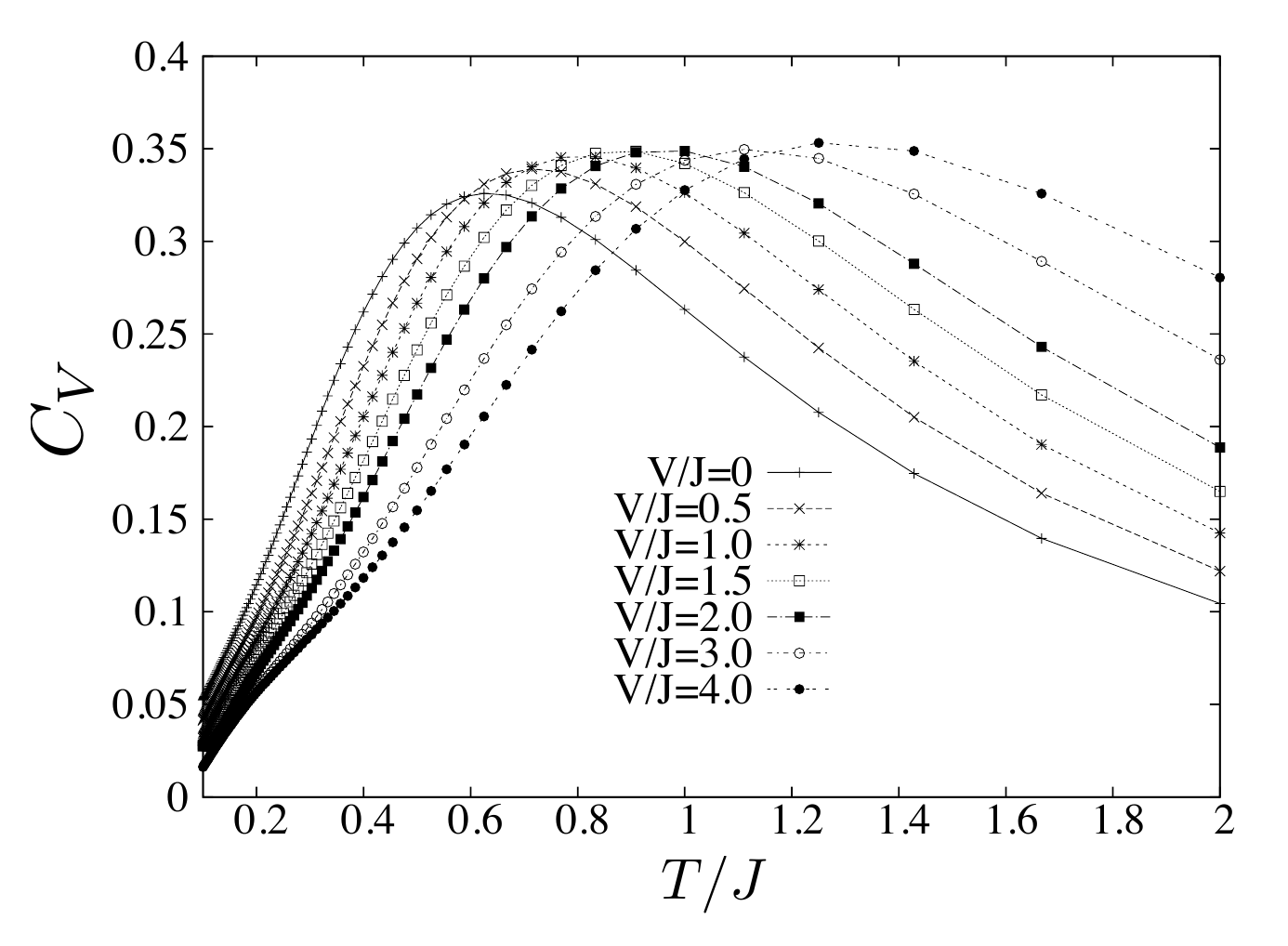}}
\caption{Specific heat of the hard-core extended Bose-Hubbard model on 34 sites for repulsive nearest neighbor interaction $V=0,0.5,1.0,1.5,2.0,3.0,4.0$.  The qualitative behavior of the low temperature specific heat changes as $V$ becomes larger than $2J$ because the system transitions from a gapless superfluid phase into a gapped insulating phase.}
\label{wall:fig:BHM}
\end{figure}

\section*{Acknowledgments}
We acknowledge useful discussions with Juan Jos\'{e} Garc\'{i}a Ripoll and Miles Stoudenmire.  This work was supported by the National Science Foundation under Grant PHY-0903457.   MLW thanks the Boulder Summer School for Condensed Matter for stimulating discussions.\\

\section*{Keywords}
`Matrix Product State'; `Bose-Hubbard Model';  `METTS'; `density matrix renormalization group';

\bibliographystyle{ws-rv-van}
\bibliography{wall.bib}

\begin{thebibliography}{28}
\providecommand{\natexlab}[1]{#1}
\providecommand{\url}[1]{\texttt{#1}}
\expandafter\ifx\csname urlstyle\endcsname\relax
  \providecommand{\doi}[1]{doi: #1}\else
  \providecommand{\doi}{doi: \begingroup \urlstyle{rm}\Url}\fi

\bibitem{white_92}
S.~R. White, Density matrix formulation for quantum renormalization groups,
  \emph{Phys. Rev. Lett.} {\bf 69}\penalty0 (19), \penalty0 2863--2866 (Nov,
  1992).
\newblock \doi{10.1103/PhysRevLett.69.2863}.

\bibitem{klumper_schadschneider_91}
A.~Klumper, A.~Schadschneider, and J.~Zittartz, Equivalence and solution of
  anisotropic spin-1 models and generalized t-j fermion models in one
  dimension, \emph{Journal of Physics A: Mathematical and General}. {\bf
  24}\penalty0 (16), \penalty0 L955,  (1991).
\newblock URL \url{http://stacks.iop.org/0305-4470/24/i=16/a=012}.

\bibitem{fannes_nachtergaele_92}
M.~Fannes, B.~Nachtergaele, and R.~F. Werner, Finitely correlated states on
  quantum spin chains, \emph{Commun. Math. Phys.} {\bf 144}, \penalty0
  443--490,  (1992).

\bibitem{oestlund_rommer_95}
S.~\"Ostlund and S.~Rommer, Thermodynamic limit of density matrix
  renormalization, \emph{Phys. Rev. Lett.} {\bf 75}\penalty0 (19), \penalty0
  3537--3540 (Nov, 1995).
\newblock \doi{10.1103/PhysRevLett.75.3537}.

\bibitem{oestlund_rommer_97}
S.~Rommer and S.~\"Ostlund, Class of ansatz wave functions for one-dimensional
  spin systems and their relation to the density matrix renormalization group,
  \emph{Phys. Rev. B}. {\bf 55}\penalty0 (4), \penalty0 2164--2181 (Jan, 1997).
\newblock \doi{10.1103/PhysRevB.55.2164}.

\bibitem{vidal_04}
G.~Vidal, Efficient simulation of one-dimensional quantum many-body systems,
  \emph{Phys. Rev. Lett.} {\bf 93}\penalty0 (4), \penalty0 040502 (Jul, 2004).
\newblock \doi{10.1103/PhysRevLett.93.040502}.

\bibitem{daley_kollath_04}
A.~J. Daley, C.~Kollath, U.~Schollwöck, and G.~Vidal, Time-dependent
  density-matrix renormalization-group using adaptive effective hilbert spaces,
  \emph{Journal of Statistical Mechanics: Theory and Experiment}. {\bf
  2004}\penalty0 (04), \penalty0 P04005,  (2004).
\newblock URL \url{http://stacks.iop.org/1742-5468/2004/i=04/a=P04005}.

\bibitem{verstraete_porras_04}
F.~Verstraete, D.~Porras, and J.~I. Cirac, Density matrix renormalization group
  and periodic boundary conditions: A quantum information perspective,
  \emph{Phys. Rev. Lett.} {\bf 93}\penalty0 (22), \penalty0 227205 (Nov, 2004).
\newblock \doi{10.1103/PhysRevLett.93.227205}.

\bibitem{verstraete_garciaripoll_04}
F.~Verstraete, J.~J. Garc\'\i{}a-Ripoll, and J.~I. Cirac, Matrix product
  density operators: Simulation of finite-temperature and dissipative systems,
  \emph{Phys. Rev. Lett.} {\bf 93}\penalty0 (20), \penalty0 207204 (Nov, 2004).
\newblock \doi{10.1103/PhysRevLett.93.207204}.

\bibitem{zwolak_vidal_04}
M.~Zwolak and G.~Vidal, Mixed-state dynamics in one-dimensional quantum lattice
  systems: A time-dependent superoperator renormalization algorithm,
  \emph{Phys. Rev. Lett.} {\bf 93}\penalty0 (20), \penalty0 207205 (Nov, 2004).
\newblock \doi{10.1103/PhysRevLett.93.207205}.

\bibitem{perezgarcia_verstraete_07}
D.~Perez-Garcia, F.~Verstraete, M.~M. WolF, and J.~I. Cirac, Matrix product
  state representations, \emph{Quantum Inf. Comput}. {\bf 7}\penalty0 (401),
  (2007).

\bibitem{verstraete_cirac_06}
F.~Verstraete and J.~I. Cirac, Matrix product states represent ground states
  faithfully, \emph{Phys. Rev. B}. {\bf 73}\penalty0 (9), \penalty0 094423
  (Mar, 2006).
\newblock \doi{10.1103/PhysRevB.73.094423}.

\bibitem{pippan_white_10}
P.~Pippan, S.~R. White, and H.~G. Evertz, Efficient matrix-product state method
  for periodic boundary conditions, \emph{Phys. Rev. B}. {\bf 81}\penalty0 (8),
  \penalty0 081103 (Feb, 2010).
\newblock \doi{10.1103/PhysRevB.81.081103}.

\bibitem{pirvu_verstraete_10}
B.~Pirvu, F.~Verstraete, and G.~Vidal, Exploiting translational invariance in
  matrix product state simulations of spin chains with periodic boundary
  conditions, \emph{arXiv:1005.5195}.  (2010).

\bibitem{mcculloch_07}
I.~P. McCulloch, From density-matrix renormalization group to matrix product
  states, \emph{Journal of Statistical Mechanics: Theory and Experiment}. {\bf
  2007}\penalty0 (10), \penalty0 P10014,  (2007).
\newblock URL \url{http://stacks.iop.org/1742-5468/2007/i=10/a=P10014}.

\bibitem{pirvu_murg_10}
B.~Pirvu, V.~Murg, J.~I. Cirac, and F.~Verstraete, Matrix product operator
  representations, \emph{New Journal of Physics}. {\bf 12}\penalty0 (2),
  \penalty0 025012,  (2010).
\newblock URL \url{http://stacks.iop.org/1367-2630/12/i=2/a=025012}.

\bibitem{stoudenmire_white_10}
E.~M. Stoudenmire and S.~R. White, Minimally entangled typical thermal state
  algorithms, \emph{New Journal of Physics}. {\bf 12}\penalty0 (5), \penalty0
  055026,  (2010).
\newblock URL \url{http://stacks.iop.org/1367-2630/12/i=5/a=055026}.

\bibitem{garciaripoll_06}
J.~J. García-Ripoll, Time evolution of matrix product states, \emph{New
  Journal of Physics}. {\bf 8}\penalty0 (12), \penalty0 305,  (2006).
\newblock URL \url{http://stacks.iop.org/1367-2630/8/i=12/a=305}.

\bibitem{feiguin_white_05}
A.~E. Feiguin and S.~R. White, Finite-temperature density matrix
  renormalization using an enlarged hilbert space, \emph{Phys. Rev. B}. {\bf
  72}\penalty0 (22), \penalty0 220401 (Dec, 2005).
\newblock \doi{10.1103/PhysRevB.72.220401}.

\bibitem{nielsen_chuang_book_00}
M.~Nielsen and I.~Chuang, \emph{Quantum Computation and Quantum Information}.
  (Cambridge University Press, Cambridge, 2000).

\bibitem{white_09}
S.~R. White, Minimally entangled typical quantum states at finite temperature,
  \emph{Phys. Rev. Lett.} {\bf 102}\penalty0 (19), \penalty0 190601 (May,
  2009).
\newblock \doi{10.1103/PhysRevLett.102.190601}.

\bibitem{vidal_03}
G.~Vidal, Efficient classical simulation of slightly entangled quantum
  computations, \emph{Phys. Rev. Lett.} {\bf 91}\penalty0 (14), \penalty0
  147902 (Oct, 2003).
\newblock \doi{10.1103/PhysRevLett.91.147902}.

\bibitem{feiguin_fiete_10}
A.~E. Feiguin and G.~A. Fiete, Spectral properties of a spin-incoherent
  luttinger liquid, \emph{Phys. Rev. B}. {\bf 81}\penalty0 (7), \penalty0
  075108 (Feb, 2010).
\newblock \doi{10.1103/PhysRevB.81.075108}.

\bibitem{Liang_Pang_94}
S.~Liang and H.~Pang, Approximate diagonalization using the density matrix
  renormalization-group method: A two-dimensional-systems perspective,
  \emph{Phys. Rev. B}. {\bf 49}\penalty0 (13), \penalty0 9214--9217 (Apr,
  1994).
\newblock \doi{10.1103/PhysRevB.49.9214}.

\bibitem{white_scalapino_09}
S.~R. White and D.~J. Scalapino, Pairing on striped $ t- t'{} -j $ lattices,
  \emph{Phys. Rev. B}. {\bf 79}\penalty0 (22), \penalty0 220504 (Jun, 2009).
\newblock \doi{10.1103/PhysRevB.79.220504}.

\bibitem{verstraete_cirac_04}
F.~Verstraete and J.~I. Cirac, Renormalization algorithms for quantum many-body
  systems in two and higher dimensions, \emph{arXiv:cond-mat/0407066v1}.
  (2004).

\bibitem{schuch_wolf_08}
N.~Schuch, M.~M. Wolf, F.~Verstraete, and J.~I. Cirac, Simulation of quantum
  many-body systems with strings of operators and monte carlo tensor
  contractions, \emph{Phys. Rev. Lett.} {\bf 100}\penalty0 (4), \penalty0
  040501 (Jan, 2008).
\newblock \doi{10.1103/PhysRevLett.100.040501}.

\bibitem{evenbly_vidal_09}
G.~Evenbly and G.~Vidal, Entanglement renormalization in two spatial
  dimensions, \emph{Phys. Rev. Lett.} {\bf 102}\penalty0 (18), \penalty0 180406
  (May, 2009).
\newblock \doi{10.1103/PhysRevLett.102.180406}.

\end{thebibliography}

\end{document}